\documentstyle[epsfig,twocolumn]{mn}

\title{Iron line profiles from black hole accretion discs with spiral
velocity structure}

\author[Sean A. Hartnoll and Eric G. Blackman]
{{Sean A. Hartnoll$^1$ and Eric G. Blackman$^2$} \\
{$^1$ DAMTP, Centre for Mathematical Sciences, Wilberforce Road, Cambridge CB3 0WA, UK.}\\
{$^2$ Department of Physics \& Astronomy, and 
Laboratory for Laser Energetics, University of Rochester,
Rochester NY 14627, USA.}}

\date{Submitted version}

\pubyear{2001}

\def\be{\begin{equation}}
\def\ee{\end{equation}}
\def\vn{{\bf \hat{n}}}
\def\vc{{\bf \hat{c}}}
\def\vr{{\bf r}}
\def\e{\epsilon}

\begin{document}

\maketitle

\begin{abstract}
We calculate the iron line profiles from accretion discs 
with spiral velocity structures 
around Schwarzschild black holes.
We find that quasi-periodic bumps appear in the 
the profiles, thereby providing a 
test for spiral wave patterns. This study is motivated by recent
work showing that spiral density waves
can result from MHD instabilities even in non-self-gravitating
discs, and by improved spectral resolution of forthcoming X-ray 
missions.

\end{abstract}


\section{Introduction}

Active galactic nuclei (AGN) and some X-ray binaries are thought to be
powered by accretion onto a central black hole (e.g. Rees 1984).
A fluorescent iron K$\alpha$ emission line with rest energy $6.4$ keV
is observed in the X-ray spectrum of many Seyfert 
galaxies and some X-ray binaries. The line profile is shaped by
D\"oppler and gravitational redshifts (Fabian {\it et al.} 1989)
and therefore by the geometry and dynamics
in the innermost regions of the accretion engine (e.g. Fabian {\it et al.}
2000 for a review).

The simplest plausible geometry to explain the observed profiles
is a geometrically thin, optically 
thick disc with an X-ray corona above and below the disc.
Some of the X-rays emanate directly to the observer,
and others are reprocessed by the disc, inducing the iron line.
General relativity is needed to calculate the redshifts
and photon trajectories. This has been done for both
Schwarzschild (Fabian {\it et al.} 1989) and Kerr (Laor 1991)
black holes. The resulting profiles have been
successful in reproducing many features of the
experimental data (e.g. Tanaka {\it et al.} 1995; Nandra {\it et al.} 1997).

The primary motivation for the present work is that 
MHD instabilities can induce spiral density waves in 
non-self-gravitating accretion discs
(Caunt and Tagger 2001, Hawley 2001, Tagger and Pellat 1999).
It is therefore interesting to see whether this spiral 
structure has a significant effect on iron line profiles. 

The effect of spiral structure on line profiles has been considered before in the context of
tidally induced spiral shocks in close binary systems (e.g. Chakrabarti and Wiita 1993,
Steeghs and Stehle 1999). The non-axisymmetry of the situation was emphasised by
Sanbuichi, Fukue and Kojima (1994), who also considered special relativistic effects.
For a review of numerical work on spiral shocks in close binaries see Matsuda {\it et al.} (2000).

This work differs from previous considerations of spiral structure 
because the context is
iron lines from AGN and therefore we use general relativity. 
For non-self-gravitating discs, and discs not in binaries,
spiral structure is still possible, but requires MHD instabilities
(Caunt and Tagger 2001, Hawley 2001, Tagger and Pellat 1999). 
The changes to the profile come purely from the
spiral {\it velocity} structure and the corresponding modified redshifts.

We find that the spiral velocity structure does have a significant 
effect on the profiles. Most notably
by introducing multiple quasi-periodic 
sub-peaks into the profile, particularly visible
for tightly packed spirals. For
reasonable values for the parameters of the spiral structure, 
the effects on the line 
profiles require better spectral resolution
than is currently available (e.g. Nandra {\it et al.} 1997), but 
will likely be discernable with Astro E-2. 
We can therefore make predictions.

Like warping (see Bachev  1999, Hartnoll and Blackman 2000)
spiral structure induces non-axisymmetric effects on the line profiles,
though the tighter the winding of the spiral the less the observed
effect.  Nevertheless, one immediate
effect of non-axisymmetry is a time dependence in the line profile of rotating discs.
Spiral shocks were used in this way by Chakrabarti and Wiita (1994) to
study variable broad optical emission lines in AGN. In fact a 
time dependence in the iron X-ray 
line is also observed (e.g. Iwasawa {\it et al.} 1996) and has been
argued to be 
problematic for the standard disc models (e.g. Weaver and Yaqoob 1998, Sulentic {\it et al.} 1998).
Although the issue remains unresolved, some kind of non-axisymmetry
of the disc may be playing a role.

It is straightforward to adapt the standard calculations of line profiles to include spiral structure.
The main difference is that the azimuthal velocity 
of the gas is no longer Keplerian and the radial velocity is nonzero.
Representative expressions for the azimuthal
and radial velocities are given in \S 2. In \S 3 we
list the formulae needed to compute the profiles, in order that our
paper be self-contained.
In particular the redshift formula needs to be generalised for non-Keplerian
rotation.
\S 4 contains sample profiles and explanation of the new features. \S 5 contains a discussion of
the results.

\section{Spiral waves in accretion discs}

We use a simple representation of spiral structure in the gas velocity, motivated by the
numerical results of Caunt and Tagger (2001)
\be
v_r = r^{-1/2} \e_r  e^{-(r-r_{in})/r_0} \cos(m \phi - k r),
\ee
\be
v_{\phi} = r^{-1/2} \left( 1 + \e_{\phi}  e^{-(r-r_{in})/r_0} \cos(m \phi - k r+ \Delta) )\right),
\ee
where $r_0$ is a constant parametrising the fall-off with radius $r$, 
$m$ is the
azimuthal wavenumber, $k$ is larger for more tightly packed spirals, $\e$ is the size of
the perturbation, $\Delta$ allows the velocities to be out of phase.
So $(\e_r,\e_{\phi},m,k,r_0,\Delta)$ is the parameter space of discs. 
We then combine these with the angle of disc inclination to the 
observerm, and the azimuthal viewing angle of the observer. 
See Fig. 1 for an example disc.
Note that in this work we consider spiral waves which are stationary 
over the observation duration.  There could in addition be
a time dependent phase in eqs. (1) and (2) to account for
a rapid spiral wave propagation.

\section{Calculation of the line profile}

\subsection{Total flux}

The total flux observed at frequency $\nu$ is given by
\be
F(\nu) \propto \int \delta (\nu - \nu_0(\vr)) f(\vr,\nu) d^2x ,
\ee
where the integral is over the disc,
$\nu_0(\vr)$ is the observed frequency
of flux emitted from $\vr$ and $f(\vr,\nu)$ is given by
\be
f(\vr,\nu) = \left[ \frac{\nu}{\nu_e} \right]^3
f_{out}(\vr,\vc(\vr,\vn)),
\ee
where $\left[\frac{\nu}{\nu_e}\right]^3$ is the relativistic correction
to the flux, with the frequency of emission,
$\nu_e = 6.4 {\mathrm{keV}}$ for the flourescent iron line. The function
$f_{out}(\vr,\vc)$ is the flux reprocessed and re-emitted
at $\vr$ in the direction $\vc$, defined such that emitted
photons in this direction from the disc eventually 
reach the observer in the Schwarzschild metric.
The asymptotic direction to the observer is $\vn$.  

The range of emitting region we consider is
$6 = r_{in} \leq r \leq r_{out} = 60$ and $0 \leq \phi < 2 \pi$,
in units with $c=G=M=1$.
The lower bound is the innermost stable orbit of the Schwarzschild metric
and the upper bound is somewhat arbitrary, but unimportant, as the objective
is to compare our results here 
with profiles from discs with axisymmetric density
structures that have well known dependences on outer radius (e.g. Fabian
{\it et al.} 1989).  Moreover, most of the interesting effects will
occur in the inner regions. 
The area element has a general relativistic correction, so
\be
d^2x = \frac{dr}{\sqrt{1-\frac{2}{r}}}r dr d\phi .
\ee

\subsection{Photon paths: direction of emission}

We use paths in the Schwarzschild metric, approximated by a first
order perturbation.
In this approximation, the initial direction $\vc$ of the photon at
$\vr$ reaching the observer, (i.e. in asymptotic direction
$\vn$) is (Hartnoll and Blackman 2001)
\be
{\bf c}(\vr) = - \frac{d \vr_{ph}}{d\Phi} (\Phi = \Phi_0) ,
\ee
where $\Phi$ parametrises the photon path and
\be
\vr_{ph}(\Phi) = r_{ph}(\Phi) \left[ \cos\Phi \vn +
\sin\Phi \frac{\vr - (\vr \cdot \vn) \vn}{r} \right] 
\ee
and the initial value $\Phi_0$ is given by
\be
\cos\Phi_0 = \frac{\vr \cdot \vn}{r} ,
\ee
with $\Phi_0 \in \left[ 0,\pi \right]$. Further $r_{ph}(\Phi)$ is
\be
\frac{1}{r_{ph}(\Phi)} = \frac{\sin\Phi}{2 R} +
\frac{1 + C\cos\Phi + \cos^2\Phi}{4 R^2} +
{\mathcal{O}} (\frac{1}{R^3}) ,
\ee
where
\be
C = -2 ,
\ee
\be
R = \frac{r \sin\Phi_0}{4} + \frac{r}{4}
\sqrt{\sin^2\Phi_0 + \frac{4}{r} (1-\cos\Phi_0)^2 } .
\ee
This perturbative approach is valid throughout most of the disc and
sufficient for our purposes. Corrections to the photon paths due
to strong curvature in the innermost regions are most significant
for very high inclination angle profiles (Matt, Perola and Stella 1993).

\subsection{Frequency shifts}

The observed frequency of radiation from $\vr$ in the Schwarzschild
metric is (e.g. Hartnoll and Blackman 2001)
\be
\nu_0(\vr) = \frac{\nu_e \sqrt{1-\frac{2}{r}
-\frac{u_r^2(\vr)}{1-\frac{2}{r}} -u_{\phi}^2(\vr)}}
{1 - \frac{\hat{c}_r(\vr) u_r(\vr)}{1-\frac{2}{r}}
 - \hat{c}_{\phi}(\vr) u_{\phi}(\vr)} .
\ee
Here ${\mathbf{u}}(\vr)$ is the velocity of the emitting material at $\vr$, given
in the previous section. As noted in Hartnoll and Blackman (2001), in this context $\vc$
needs to be normalised so that $\frac{\hat{c}_r(\vr) \hat{c}_r(\vr)}{1-\frac{2}{r}}
+ \hat{c}_{\theta}(\vr) \hat{c}_{\theta}(\vr) + \hat{c}_{\phi}(\vr) \hat{c}_{\phi}(\vr)
= \frac{1}{1-\frac{2}{r}}$. Fig. 2 shows the redshifts for a sample disc.

\subsection{Reprocessed flux}

We need to calculate $f_{out}(\vr,\vc)$. We use the standard iron line model
(Fabian {\it et al.} 1989) in which there is assumed to be a corona above the central
region producing hard X-rays which are reprocessed by the disc. The corona is
modelled by a source which is at height $h_s$ directly above the centre
of the disc. So the flux impinging on the disc at $\vr$ is
\be
f_{in}(\vr) \propto \frac{h_s}{(h_s^2 + r^2)^{q/2}} .
\ee
We will take $q=3$ (i.e. $1/r^2$ falloff from source) and $h_e = 10$. The main objective is to
study the effect of spatial inhomogenity on the line profiles, so the exact values of these parameters
are not important so long as they allow us to compare the results with standard line profiles.
Assuming the re-emission is isotropic, we have
\be
f_{out}(\vr,\vc) \propto f_{in}(\vr) {\bf e_z} \cdot \vc ,
\ee
where ${\bf e_z}$ is the unit normal to the disc. Note that the varying density
will affect the optical depth but not the total power coming from a given radius as long as the optical depth is $>> 1$.
We do not incorporate the vertical density structure of
the disc in calculating our line profiles.

\section{Discussion of Line profiles}

The region of the parameter space investigated in this section is 
qualitatively guided by
recent numerical results from simulations of spiral structure in non-self-gravitating MHD
accretion discs (Caunt and Tagger 2001, Hawley 2001).

Fig. 3 shows profiles from discs at $30$ degrees inclination for various values of $k$
and $r_0$. The spiral velocity structure results in the presence of small peaks across
the profile. As expected, 
increasing $k$ increases the frequency of the the peaks.
At lower $r_0$, the spiral structure is more confined to the
inner regions. Note that $\e=\e_r=\e_{\phi}$ needs to be changed with $r_0$ 
to ensure that the overall normalisation is comparable.
A concentration of structure at small $r$ reveals itself
through the prevalence of peaks or bumps 
appearing in the more redshifted or blueshifted regions of the profile.

Fig. 4 shows profiles from discs at $15$ degrees inclination for various 
 $k$ and a 
profile with no spiral structure for comparison. A larger range of $k$ is 
considered in comparison to 
Fig. 3. Again the correlation between number of peaks and $k$ is evident.

Fig. 5 shows profiles from discs at an inclination angle of $60$ degrees. The profiles from
discs at high inclination angles are naturally more spread out. The effects of $k$ and $r_0$
are as before. It is particularly noticeable in these profiles that the sub-peaks occur only
outside of the standard double peaks. This is due to the falloff with $r$; the spirals are more
prominent in the inner regions of the disc, which are more D\"oppler shifted.
The atypical blueshifted region of the $k=1.2$, $r_0=15$ graph is also interesting,
the sub-peaks evolve into a step-like form.

All discs considered above have $m=1$, $\Delta=0$, $\e_r = \e_{\phi}$ fixed given $r_0$,
and the observer is at the same azimuthal angle in all cases.
Varying $\e$ with $r_0$ fixed has the obvious effect of increasing the relative magnitude of
the perturbations to the usual line profile from discs without spiral structure. Varying $m$,
$\Delta$ or the azimuthal angle of the observer does not appear to have a significant qualitative
effect on the profiles.

\section{Conclusion}

The most interesting overall feature in our line profiles is
the presence of multiple, quasi-periodic 
sub-peaks in the most red or blue-shifted regions,
indicating spiral structure in the inner disc.
The profiles can be compared with data to
infer a constraint on dynamical models predicting spiral structure. 
Figs. 3, 4 and 5 show what sort of values of $\e$, for a given $r_0$, 
are consistent with present data.

Though not easily constrained with current instruments, this quasi-periodic
bumpiness should be resolvable with Astro E-2, which carries the
X-ray Spectrometer (XRS) of spectral resolution $\Delta E/E \sim 0.002$
in the $6.4$keV range (Kelley {\it et al.} 1999) which should be sufficient
for a wide range of $k$.
For very low $k$, there are fewer sub-peaks, and their exact
shape and position depend on the variables $m,\Delta,\frac{\e_r}{\e_{\phi}}$
and the azimuthal angle of the observer. 

\section*{Acknowledgements}
EGB acknowledges partial support from DOE grant DE-FG02-00ER54600.

\vfill
\eject


\noindent\textbf{Figure 1:} The spiral structure of the radial velocity in a disc with $m=1$,
$k=1.2$, $r_0=15$.

\medskip


\noindent\textbf{Figure 2:} The redshifts of the disc of Fig. 1, viewed at an inclination
angle of $30$ degrees. Darker shading indicates higher redshift. The disc is rotating
anticlockwise, the observer is situated to the right.


\medskip

\noindent\textbf{Figure 3:} Profiles from discs at inclination angle $30$ degrees with $m=1$,
$\Delta=0$. Left column: $k=0.4$. Right column: $k=1.2$. Top row: $r_0 = 5, \e = 0.3$.
Bottom row: $r_0 = 15, \e = 0.2$.

\medskip

\begin{figure}
\epsfig{file=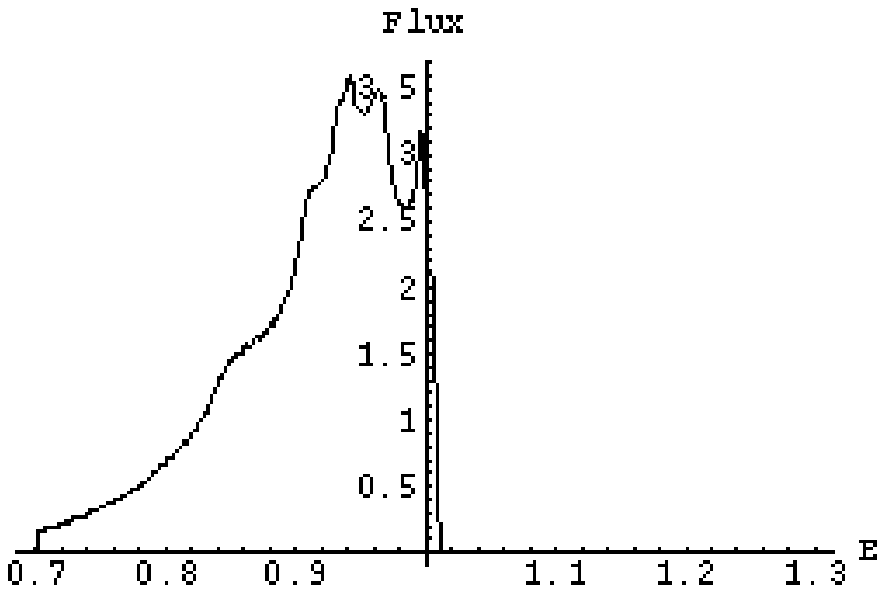,width=1.8in}\epsfig{file=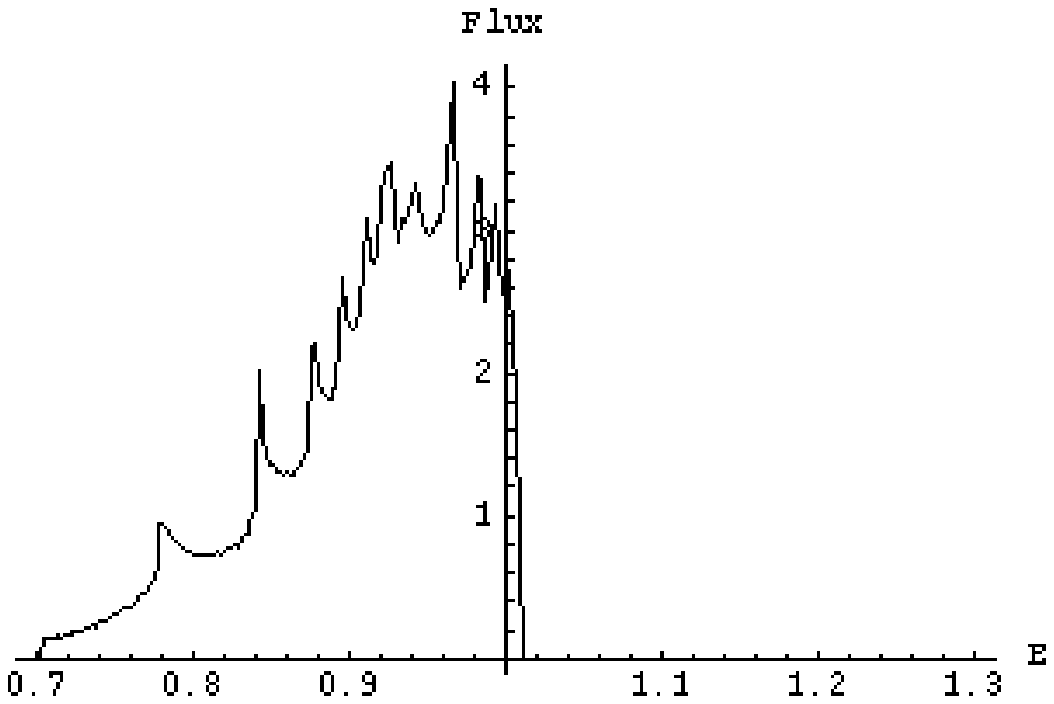,width=1.8in}\\
\epsfig{file=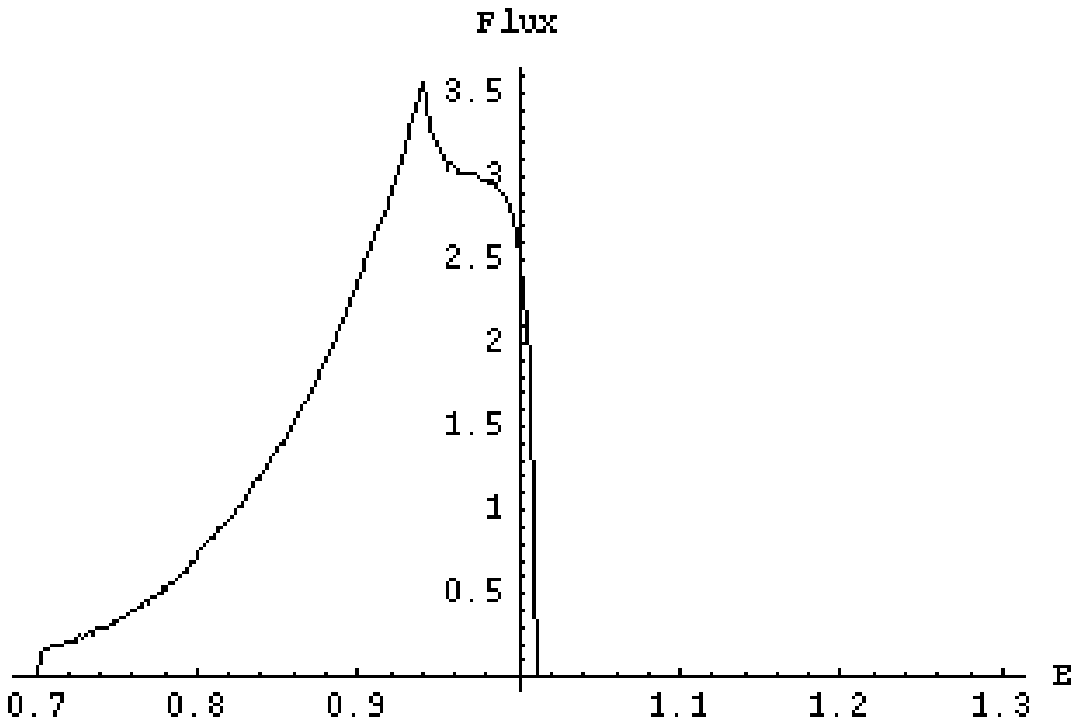,width=1.8in}\epsfig{file=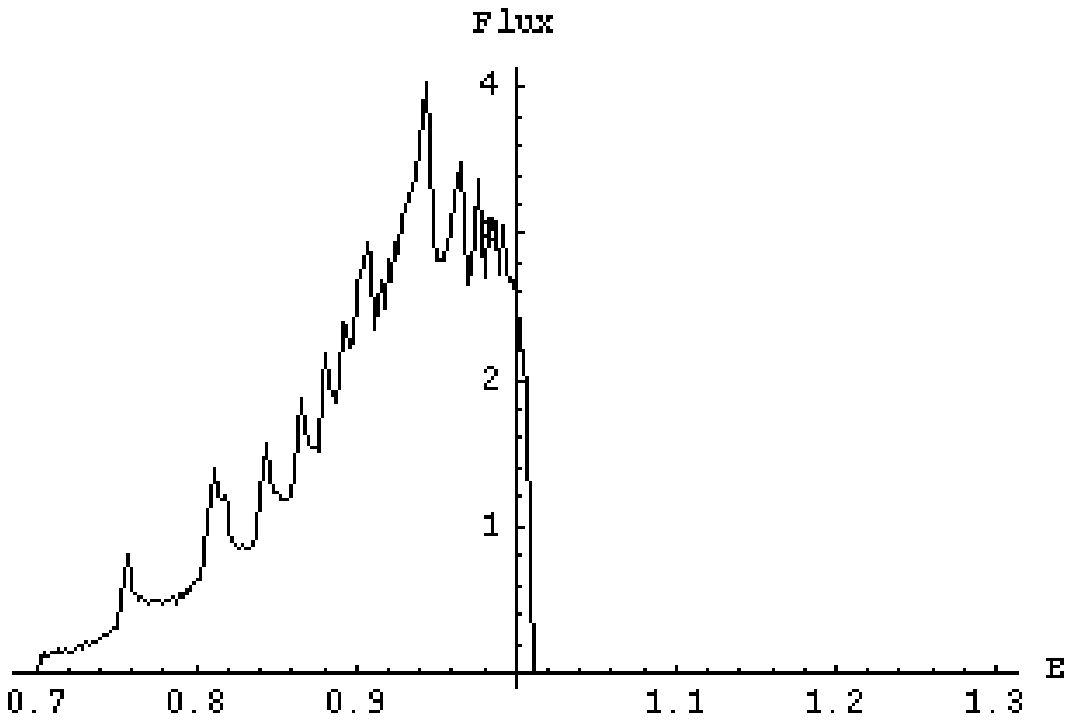,width=1.8in}\\

\noindent\textbf{Figure 4:} Clockwise from top left: The first three are profiles from
discs at inclination angle $15$ degrees with $m=1$, $\Delta=0$,
$r_0=15$,$\e=0.2$. From left to right: $k=0.4$, $k=1.2$, $k=2.0$. The profile
at the bottom left is from a disc with no spiral structure, for comparison.

\end{figure}

\begin{figure}
\epsfig{file=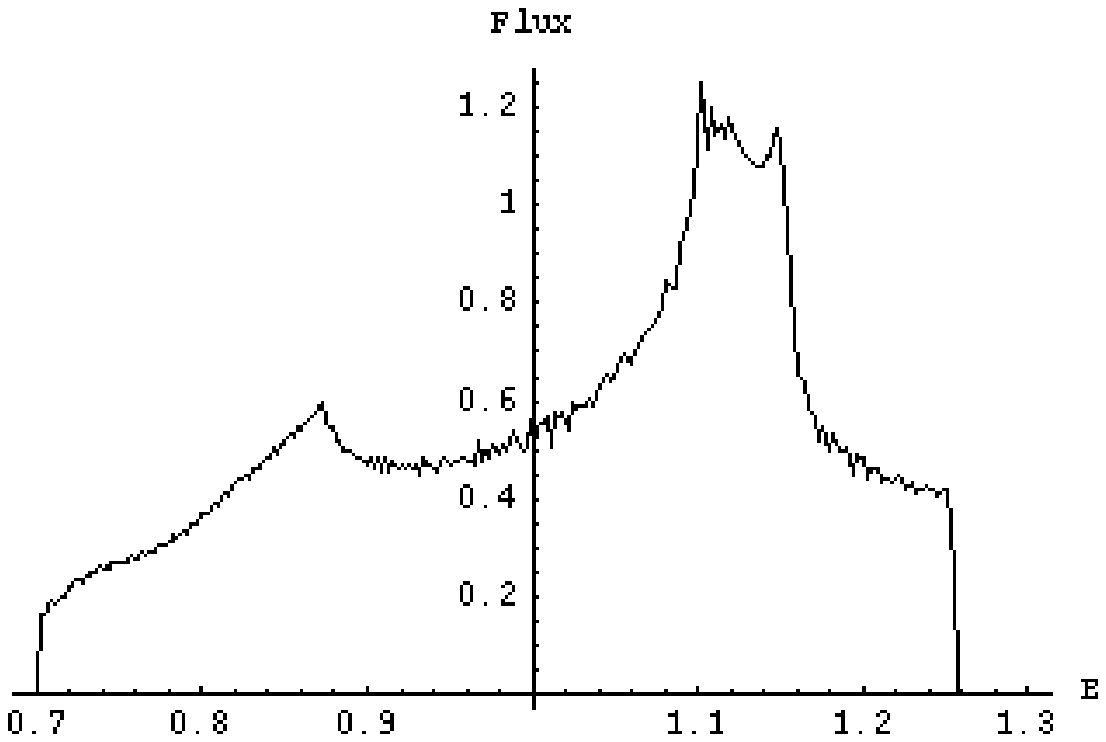,width=1.8in}\epsfig{file=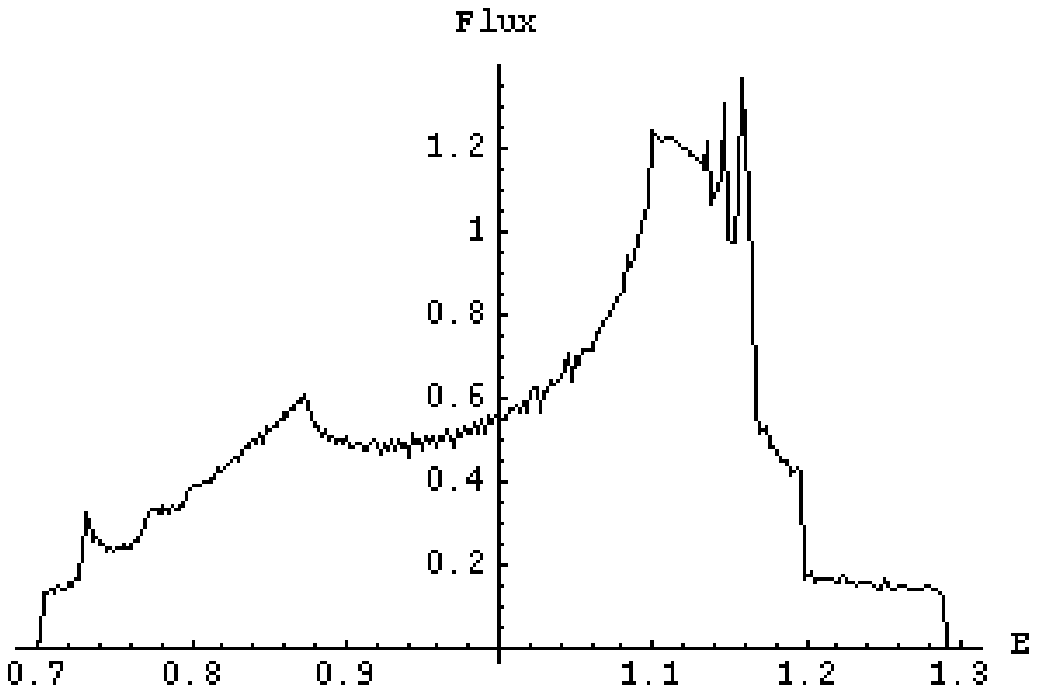,width=1.8in}\\
\epsfig{file=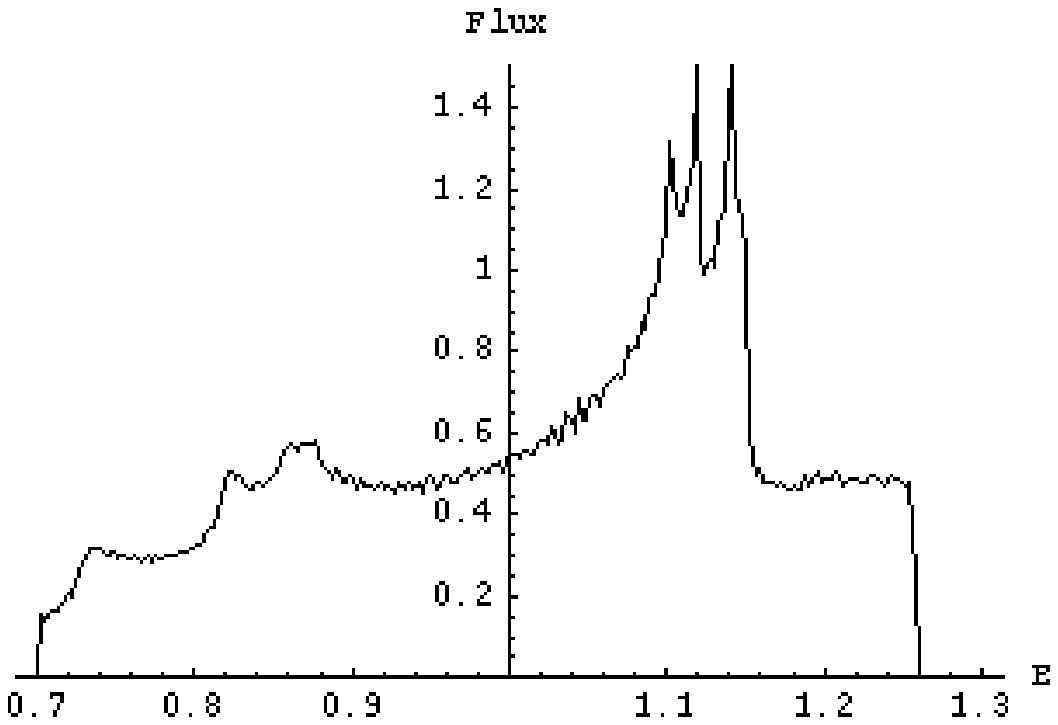,width=1.8in}\epsfig{file=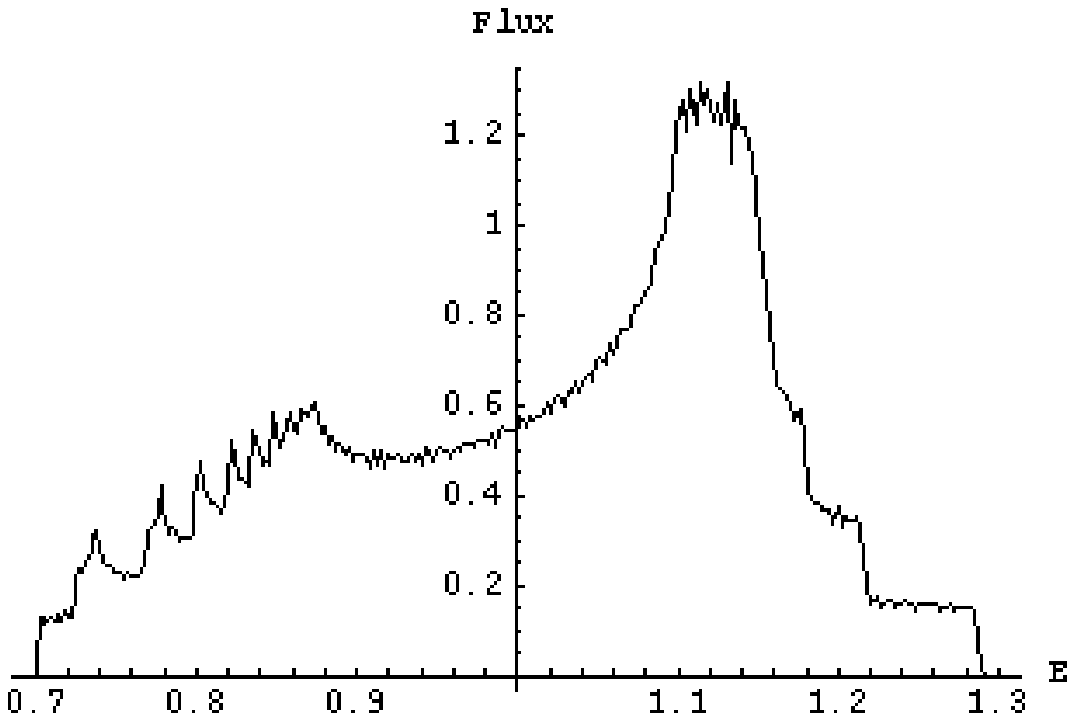,width=1.8in}\\

\noindent\textbf{Figure 5:} As for Fig. 3 but at inclination angle of $60$ degrees.
Left column: $k=0.4$. Right column: $k=1.2$. Top row: $r_0 = 5, \e = 0.3$. Bottom row:
$r_0 = 15, \e = 0.2$.

\end{figure}


\begin{thebibliography}{}

\bibitem{} Bachev R., 1999, A\& A, 348, 71

\bibitem{} Caunt S.E. \& Tagger M., 2001, A\&A, 367, 1095

\bibitem{} Chakrabarti S.K. \& Wiita P.J., 1993, A\&A, 271, 216

\bibitem{} Chakrabarti S.K. \& Wiita P.J., 1994, ApJ, 434, 518

\bibitem{} Fabian A.C., Rees M.J., Stella L. \& White N.E., 1989, MNRAS, 238, 729

\bibitem{} Fabian A.C., Iwasawa K., Reynolds C.S. \& Young A.J., 2000, PASP, 112, 1145.

\bibitem{} Hartnoll S.A. \& Blackman E.G., 2000, MNRAS, 317, 880

\bibitem{} Hartnoll S.A. \& Blackman E.G., 2001, MNRAS, 324, 257

\bibitem{} Hawley J.F., 2001, ApJ, 554, 534

\bibitem{} Iwasawa K., {\it et al.}, 1996, MNRAS, 282, 1038

\bibitem{} Kelley R.L. {\it et al.} 1999, Proc. SPIE, 3765, 114 

\bibitem{} Laor A., 1991, ApJ, 376, 90

\bibitem{} Matsuda T., Makita M., Fujiwara H., Nagae T., Haraguchi K., Hayashi E., Boffin H.M.J,
2000, Ap\&SS, 274, 259

\bibitem{} Matt G., Perola G.C. \& Stella L., 1993, A\&A, 267, 643

\bibitem{} Nandra K., George R.F., Mushotzky R.F., Turner T.J. \& Yaqoob T., 1997, ApJ, 477, 602

\bibitem{} Rees M.J., 1984, ARA\&A, 22, 417

\bibitem{} Sanbuichi K., Fukue J., Kojima Y., 1994, PASJ, 46, 605

\bibitem{} Steeghs D., Stehle R., 1999, MNRAS, 307, 99

\bibitem{} Sulentic J.W., Marziani P., Calvani M., 1998, ApJ, 497, L65

\bibitem{} Tagger M. \& Pellat R., 1999, A\&A, 349, 1003

\bibitem{} Tanaka Y. {\it et al.}, 1995, Nature, 375, 659

\bibitem{} Weaver K.A. \& Yaqoo T., 1998, ApJ, 502, L139

\end{thebibliography}
\end{document}